\documentclass{phc-proc4-auth}

\usepackage{graphicx}

\begin{document}

\begin{frontmatter}



\title{Metallic charge stripes in cuprates}


\author{J. M. Tranquada}

\address{Physics Department, Brookhaven National Laboratory, Upton, NY
11973-5000, USA}

\begin{abstract}
Some recent evidence for the existence of dynamic, metallic stripes in
the 214-family of cuprates is reviewed.  The mechanism of stripe pinning
is considered, and changes in the charge density within stripes between
the pinned and dynamic phases is discussed.  From a purely experimental
perspective, dynamic charge stripes are fully compatible with nodal
``quasiparticles'' and other electronic properties common to all
superconducting cuprates.
\end{abstract}

\begin{keyword}
stripes \sep neutron scattering \sep 214 cuprates
\PACS 74.72.Dn \sep 71.45.Lr \sep 75.30.Fv
\end{keyword}
\end{frontmatter}

\section{Introduction}
\label{}

It is currently fashionable to consider phases that compete with
superconductivity in the cuprates.  These may be invoked to explain the
pseudogap, or to characterize the state within a vortex core.  Charge and
spin stripes represent a specific type of order that has been
experimentally observed to compete with superconductivity
\cite{tran97a,ichi00}, to be pinned by vortices \cite{lake02,khay02}, and
to be pinned by Zn impurities
\cite{hiro98}.

It has been speculated that dynamic stripes may underly the
superconducting state, and, indeed, a plausible mechanism for
superconductivity based on stripe correlations has been proposed
\cite{emer97,carl03}.  In this paper, I review some of the evidence from
cuprates in the 214 family that dynamic stripes exist and are metallic. 
The cause of stripe pinning and anomalies in the pinned state are also
discussed.

\section{Evidence for the stripe-liquid phase}

While ordered charge stripes can coexist with superconductivity
\cite{tran97a}, stripe order is bad for superconductivity \cite{ichi00}. 
If stripes are to have a positive relevance to superconductivity in the
cuprates, then they must be able to exist in a dynamic form.  It has
recently been shown that the insulating diagonal stripes of the layered
nickelates can melt into a liquid phase \cite{lee02}.  The continuous
evolution of the magnetic correlations from the ordered state to the
disordered state provides direct evidence for dynamic charge stripes.

To obtain similar evidence for dynamic stripes in cuprates, inelastic
neutron scattering experiments have been performed on
La$_{1.875}$Ba$_{0.125}$CuO$_4$.  Fujita {\it et al.} \cite{fuji02} have
successfully grown a single crystal of this composition.  The structure
transforms from the low-temperature-orthorhombic (LTO) phase, common to
La$_{2-x}$Sr$_x$CuO$_4$, to the low-temperature-tetragonal (LTT)
phase on cooling through 60 K.  Superlattice peaks characteristic of
charge and spin stripe order appear below $\sim50$~K \cite{fuji02,fuji03}.

At 30 K, well within the stripe-ordered regime, the inelastic
measurements reveal, in $\omega$-{\bf Q} space, magnetic excitations
rising steeply out of the superlattice positions (the measurements extend
up to 12 meV) \cite{fuji03}.  The scattering is fairly sharp in {\bf Q},
and the intensity, corrected for the Bose factor, is independent of
excitation energy, consistent with what one would expect for {\bf
Q}-integrated spin waves.  On warming through the stripe melting
transition and into the LTO phase, the magnetic excitations evolve
continuously.  At 65 K, the scattering peaks measured at a given energy
have shifted to a slightly smaller splitting from the antiferromagnetic
wave vecter, they have broadened in {\bf Q} width, and there is a loss
of low-energy weight in the Bose-factor-corrected intensity.  (Similar
results have been obtained on La$_{1.48}$Nd$_{0.4}$Sr$_{0.12}$CuO$_4$ by
Ito {\it et al.} \cite{ito03}.)

The incommensurate magnetic excitations in the LTO phase are clearly
connected with the magnetic order in the LTT phase.  Given that the spin
order is tied to charge-stripe order, it follows that the magnetic
excitations in the disordered state provide direct evidence for dynamic
charge stripes.  The similarity of the spin correlations observed in
La$_{1.875}$Ba$_{0.125}$CuO$_4$ with those found in
La$_{2-x}$Sr$_x$CuO$_4$ \cite{aepp97,yama98a} strongly indicates that
dynamic stripes are a common feature of the 214 cuprates.

\section{Stripe pinning mechanism}

Kivelson, Fradkin, and Emery \cite{kive98} have proposed that dynamic
stripes may be thought of in terms of an electronic liquid-crystal
model.  Within this model, each charge stripe is modeled as a
one-dimensional electron gas, and there is a competition between
instabilites towards charge-density-wave (CDW) correlations and
superconductivity within the stripes.  Fluctuations of the stripes favor
two-dimensional superconductivity, but pinning of straight, parallel
stripes would lead to CDW order.

Hasselmann, Castro Neto, and Morais Smith \cite{hass02} have analyzed a
model for the coupling between longitudinal charge dynamics with
transverse fluctuations for a single stripe.  They find that a zig-zag
transverse lattice potential can induce a $4k_{\rm F}$ CDW state within
the stripe.  Such a transverse potential occurs within the CuO$_2$ planes
of the LTT phase due to the modulations of the positions of the in-plane
oxygens along the $c$ axis.  Thus, one might expect static stripe order
in the LTT phase to be correlated with the appearance of a CDW gap in the
optical conductivity.  ($4k_{\rm F}$ CDWs within charge stripes have also
been considered by Zaanen and Ole\'s \cite{zaan96b}.)

Dumm {\it et al.} \cite{dumm02} have measured the in-plane optical
conductivity in stripe ordered La$_{1.275}$Nd$_{0.6}$Sr$_{0.125}$CuO$_4$
vs.\ temperature.  (Related results have been reported by Tajima {\it et
al.} \cite{taji99}.)  Down to 32~K the results look very similar to those
for superconducting La$_{1.875}$Sr$_{0.125}$CuO$_4$, but at 5~K one
observes a peak at finite frequency ($\sim8$~meV).  
Although Dumm {\it et al.} \cite{dumm02} have interpreted this feature in
terms of localization effects, it also appears qualitatively consistent
with CDW behavior.

\section{Evidence that stripes are metallic}

It has been argued above that stripe correlations are common among the
214 cuprates.  One expects the existence of stripe correlations to have a
strong impact on the electronic excitations \cite{kive03}.  Thus, it
follows that characterizations of the electronic properties of
La$_{2-x}$Sr$_x$CuO$_4$ should reflect the nature of dynamic stripes.

The in-plane optical conductivity observed for La$_{2-x}$Sr$_x$CuO$_4$
\cite{dumm02,taji99,gao93} looks quite similar, in both frequency and
temperature dependence, to that measured on cuprates with higher
superconducting transition temperatures \cite{rome92}.  Angle-resolved
photoemission measurements on La$_{2-x}$Sr$_x$CuO$_4$ \cite{zhou01} reveal
a distribution of spectral weight near the Fermi surface that is very
similar to that measured on optimally-doped
Bi$_2$Sr$_2$CaCu$_2$O$_{8+\delta}$ \cite{vall00}.  In particular,
spectral weight is found near the ``nodal'' point even in stripe-ordered
La$_{1.4-x}$Nd$_{0.6}$Sr$_x$CuO$_4$ \cite{zhou01}.

The existence of metallic properties in La$_{2-x}$Sr$_x$CuO$_4$ indicates
that the dynamic charge stripes must be metallic.  The fact that the
electronic response of this material is essentially the same as that of
other cuprates, together with the idea that stripes should influence the
electronic properties, suggests that stripe correlations (or a related
form of instantaneous charge inhomogeneity) could be ubiquitous in the
superconducing cuprates.  Further evidence for stripe correlations in a
variety of cuprates is reviewed in \cite{kive03}.

\section{Mid-gap states associated with stripes}

Theoretical analyses of cuprates have predicted that the dopant-induced
states associated with stripes (or more general forms of charge
inhomogeneity) should appear within the charge-transfer gap of the parent
insulator \cite{zaan89,emer93,lore02}.  Such a concept should also apply
to the diagonal charge stripes found in La$_{2-x}$Sr$_x$NiO$_{4+\delta}$;
indeed, the schematic density of states shown in Fig.~1 has been used
recently to explain the mid-infrared peaks observed in optical
conductivity \cite{home03}.  The mid-gap states in the nickelates are
empty at low temperature \cite{zaan94}, so that these materials are
insulating.  

\begin{figure}[t]
\centerline{\includegraphics[width=2.8in]{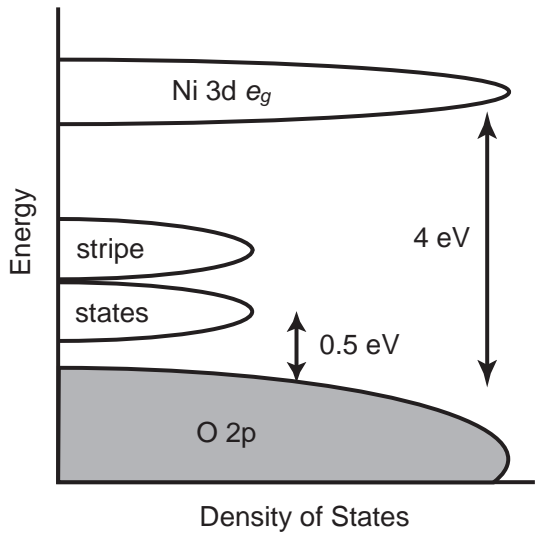}}
\medskip
\caption{Schematic diagram of the electronic density of states associated
with charge stripes in La$_{2-x}$Sr$_x$NiO$_{4+\delta}$
\protect\cite{home03}. }
\label{fg:rho} 
\end{figure}

In contrast to the nickelates, one expects that ordered vertical stripes
in the cuprates should be precisely $\frac14$-filled with electrons, as
indicated in Fig.~2.  In this case, there is electron-hole symmetry, so
that the off-diagonal conductivity should go to zero
\cite{emer00,prel01}, as observed in the Hall-effect study of
stripe-ordered La$_{1.4-x}$Nd$_{0.6}$Sr$_x$CuO$_4$ by Noda {\it et al.}
\cite{noda99}.  The anomalous vanishing of the Hall coeficient occurs
only when the stripes are ordered, as in the LTT phase; the Hall
coefficient behaves ``normally'' in the LTO phase.  To reconcile this
``normal'' behavior with the existence of dynamic stripes, it is
necessary to take account of the shift of the incommensurability with
temperature measured by inelastic neutron scattering
\cite{fuji03,ito03}.   The incommensurability is proportional to the
density of stripes.  If the density of stripes changes while the net
density of holes associated with stripes remains unchanged, then the
electron (or hole) filling of the stripes will change.  Any deviation
from the precise $\frac14$ filling of the ordered state will break
particle-hole symmetry, and hence the anomalous Hall effect will
disappear.   The variation in stripe-filling in the dynamic phase is also
consistent with frustration of the CDW instability by stripe
fluctuations \cite{kive98}.  

\begin{figure}[t]
\centerline{\includegraphics[width=2.8in]{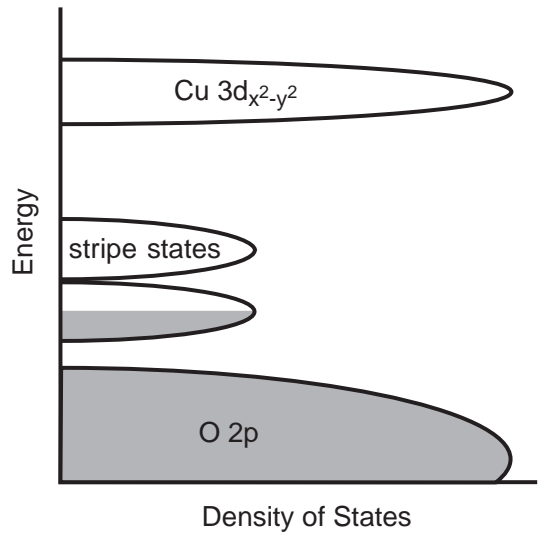}}
\medskip
\caption{Schematic diagram of the electronic density of states associated
with charge stripes in cuprates.}
\label{fg:rho} 
\end{figure}

\section{Conclusion}

We have seen that there is strong evidence that dynamic stripes exist in
La$_{2-x}$Sr$_x$CuO$_4$ and related cuprates.  In such
materials, photoemission studies show the existence of nodal
``quasiparticles'', and the optical conductivity looks similar to that of
many other superconducting cuprates.  Pinning of stripes to lattice
modulations appears to be associated with a CDW instability within the
stripes.  Changes in the electronic charge density within the charge
stripes are correlated with the differences between pinned and dynamic
stripes.

Experimentally, dynamic charge stripes are found to be compatible with
the typical electronic properties of most cuprates.  Of course, features
such as nodal quasiparticles do not arise naturally out of theoretical
models that focus on the one-dimensional nature of stripes.  Thus, there
remains a theoretical challenge to understand how the electronic
properties of cuprates can be derived from dynamic stripes.  It should be
kept in mind that a lack of theoretical understanding does not invalidate
the experimental case for metallic dynamic stripes. 

\section{Acknowledgements}

I have benefited from numerous discussions with S. A. Kivelson and E.
Fradkin, and I am especially grateful to my experimental collaborators M.
Fujita, H. Goka, and K. Yamada.
Research at Brookhaven
is  supported by the Department of Energy's (DOE) Office  of Science
under Contract No.\ DE-AC02-98CH10886.










\end{document}